\documentclass{aa}
\usepackage{graphicx}

\def\deg{$^\circ$}
 
\def\arcsec{$^{\prime\prime\,}$}
\def\arcmin{$^{\prime\,}$}

\def\a{$^{\mbox{\small a}}$}
\def\b{$^{\mbox{\small b}}$}
\def\c{$^{\mbox{\small c}}$}
\def\d{$^{\mbox{\small d}}$}
\def\e{$^{\mbox{\small e}}$}
\def\f{$^{\mbox{\small f}}$}
\def\g{$^{\mbox{\small g}}$}
\def\*{$^{*}$}

\def\nat{Nature}
\def\aap{A\&A}
\def\apjl{ApJL}
\def\apj{ApJ}
\def\apjs{ApJ Suppl.Ser.}
\def\iaucirc{IAUC}
\def\mnras{MNRAS}

\begin{document}

\sloppypar
 
\title{INTEGRAL insight into the inner parts of the Galaxy. High mass X-ray 
binaries.}
 
\author{A.Lutovinov\inst{1,2}, M.Revnivtsev\inst{2,1}, M. Gilfanov\inst{2,1},
P.Shtykovskiy\inst{1,2}, S. Molkov \inst{1,2}, R.Sunyaev\inst{1,2}}      

\offprints{lutovinov@hea.iki.rssi.ru} 

\institute{Space Research Institute, Russian Academy of Sciences,
              Profsoyuznaya 84/32, 117810 Moscow, Russia
\and
Max-Planck-Institute f\"ur Astrophysik, Karl-Schwarzschild-Str. 1, D-85740
Garching bei M\"unchen, Germany
}

\date{Received /Accepted}

        \authorrunning{Lutovinov et al.}
        \titlerunning{INTEGRAL insight into the inner parts of the Galaxy} 
        
\abstract{Since its launch INTEGRAL has devoted a significant fraction of
its observing time to the Galactic plane region. We present the results of
INTEGRAL observations of the inner spiral arms of the Galaxy (Norma, Scutum
and Sagittarius) with attention to high mass X-ray binaries. The increased
sensitivity of the survey and possibility to detect absorbed sources
significantly enlarged our sample of HMXBs in comparison with previous
studies. For some absorbed sources we present for the first time high
quality broadband (1-100 keV) energy spectra. We confirm pulsations of the
X-ray flux of IGR J16358-4726, discovered by CHANDRA, and report discovery
of pulsations with a period of $228\pm6$ sec from IGR J16465-4507. We
confirm that the Galactic high mass X-ray binary distribution is
significantly different from the distribution of low mass X-ray binaries
concentrated in the Galactic Center. A large fraction of detected high mass
X-ray binaries are absorbed sources.  
 }

   \maketitle
%
%________________________________________________________________ 

\section{Introduction}

One of the most interesting results obtained by the INTEGRAL observatory is
the discovery of a large population of highly photoabsorbed Galactic X-ray
sources. The first source discovered with INTEGRAL (IGR J16318-4848,
\cite{cour03}) was shown to have a large intrinsic absorption in the X-ray
spectrum (e.g.  \cite{murakami03}, \cite{deplaa03}, \cite{matt03},
Rev\-niv\-tsev et al. 2003a).  After this source several more were
discovered (e.g. IGR J16320-4751=AX J1631.9-4752, IGR J16358-4726, etc.)
which also have high photoabsorption, strongly exceeding that of the
interstellar medium.  It was proposed that these sources can be members of a
previously poorly known population of Galactic X-ray sources with high
intrinsic absorption (\cite{rev03a}). Based on the photometric observations
of the source IGR J16318-4848 these authors also suggested that the source
could be a binary with a giant or supergiant companion with a strong stellar
wind, that was later confirmed by precise spectroscopic observations of the
optical counterpart (\cite{chaty04}).

The analysis of properties of newly discovered absorbed sources showed that
they are likely high mass X-ray binaries (HMXBs) with early type companion
stars, and some of them contain accreting X-ray pulsars (\cite{patel04},
\cite{walter04}, \cite{lut05a}).  Remarkably these sources are confined in a
relatively small region of the sky, close to the tangent to the Norma
Galactic spiral arm.  Based on a previous study of the population of high
mass X-ray binaries in our Galaxy (Grimm et al. 2002) it was proposed
(Revnivtsev 2003) that such a concentration of absorbed HMXBs is the result
of an enhanced star formation rate in this region and is an observational
appearance of the Galactic spiral arm structure.

A similar enhanced population of HMXBs (more specifically, accreting X-ray
pulsars) in the regions of spiral arms was emphasized by Koyama et
al. (1990). Basing on the GINGA observations it was shown that in the region
of the Scutum spiral arm tangent there exists a large number of accreting
X-ray pulsars, likely variable, that was considered as an indication that
they are in binary systems with $B[e]$ companion stars. A subsequent study
of these systems with the ASCA observatory showed that many of them are
absorbed (\cite{sugizaki01}), and the comparison of the measured value $N_H$
with the interstellar Galactic absorption indicates that the observed X-ray
photoabsorption is likely intrinsic.

Previous studies of Galactic X-ray sources (e.g. Koyama et al. 1990,
Negueruela 2004) suggest that sources with significant intrinsic
photoabsorption in their spectra can be unambiguously classified as HMXBs,
while the absence of large photoabsorption does not allow a reliable
classification.

Such a difference in observational appearances of high mass and low mass
X-ray binaries can be understood if we remember that in the majority of high
mass X-ray binaries the compact objects accrete matter from the stellar wind
of a young companion star, and therefore the dense stellar wind can create
an envelope which will result in significant photoabsorption within the
binary system. The best known examples of high mass X-ray binaries in our
Galaxy are Cyg X-3, Vela X-1, GX301-2 and 4U1700-37. They all show the
strong photoabsorption in their X-ray spectra.

In the case of low mass X-ray binaries (LMXBs) the main accretion goes
through the inner Lagrangian point and therefore practically no obscuring
matter is available in such systems (except for the rare class of dippers or
accretion disk corona sources, see e.g. Kahn 1982). From this reasoning we
can expect that any sample of HMXBs selected from soft X-ray data might lack
systems with high intrinsic photoabsorption and therefore will lack a
considerable number of HMXBs. In order to construct more uniform sample of
HMXBs a hard X-ray ($>10-20$ keV) energy band survey is important.

The international gamma ray observatory INTEGRAL (\cite{win03}) is useful
for the search of absorbed sources. The telescope IBIS/ISGRI (\cite{ube03},
\cite{leb03}) has a large effective area ($\sim$1000 sq.cm.) and works in
the hard X-ray energy range ($>20$ keV), that largely excludes any influence
of the photoabsorption of X-ray sources on the detection efficiency. It has
a large field of view ($\sim 29^\circ\times 29^\circ$) and good localization
accuracy ($\sim2-3$\arcmin).

The first two years of INTEGRAL observations provided more than 20 million
seconds of data. The deepest hard X-ray ($\sim$20-60 keV) surveys to date of
the Galactic Center region -- down to $\sim$1 mCrab level (\cite{rev04a}),
the Sagittarius Arm region -- down to $\sim$1.5 mCrab level
(\cite{molkov04}) and the Galactic plane -- down to $\sim$2-3 mCrab level
(\cite{bird04}) were published. Here 1 mCrab corresponds approximately to
$\sim 1.4\times 10^{-11}$ erg s$^{-1}$ cm$^{-2}$ for a source with a
Crab-like spectrum. Much INTEGRAL data has became public and this opens the
possibility to construct a highly sensitive flux limited sample of sources
in the inner part of the Galaxy and try to make a systematic study of their
broadband X-ray properties and their distribution in the Galaxy.

In this paper we focus on the analysis of the population of high mass X-ray
binaries located in the Galactic plane from the Norma to Sagittarius spiral
arms. This Galactic region was selected because it has the best statistics
of the available INTEGRAL data.  We plan to extend this study towards
further Galactic spiral arms when more INTEGRAL data are available.  We
present the high quality broadband (1-100 keV) spectra of a number of newly
discovered absorbed objects and report the discovery of X-ray pulsations
from IGR J16465-4507.

\section{Data reduction}

For the reconstruction of the source broadband spectra we used all available
X-ray data, which include data of the INTEGRAL observatory (all public data
for revolutions up to 118 and the General Program data of the Galactic
Center observations, Obs.ID 0120213 and 0220133) as well, data of RXTE, ASCA
and XMM-Newton observatories. The data of the latter observatories were
reduced using standard tools recommended by the Guest Observer Facilities
(http://legacy.gsfc.nasa.gov).  The data of all INTEGRAL/IBIS/ISGRI
observations were processed with the method described by Revnivtsev et
al. (2004). In order to reconstruct the source spectra from IBIS/ISGRI data
we used a ratio of fluxes measured in different energy channels to the
fluxes measured by the ISGRI detector from the Crab nebula in the same
energy bands. Detailed analysis of Crab nebula observations suggests that
with the approach and software employed, a conservative estimation of
uncertainty in measurements of absolute fluxes from the sources is about
10\% and the shape of the spectrum is about 5\%. The last value was added to
the following spectral analysis as a systematic uncertainty in each energy
channel.

RXTE spectra of IGR J16318-4848 and IGR J16358-4726 were taken from the work
of Revnivtsev (2003). In this work the special treatment was done in order
to get rid of a strong contribution of the Galactic ridge emission.

\section{The sample and completeness}

The available INTEGRAL data now contain approximately $>$20 Msec of
observations, in particular $\sim>$5 Msec of the inner parts of the Galaxy
($|l|<90^\circ$).  In this paper we focus on the Galactic regions with
Galactic latitudes $|b|<5^\circ$ and Galactic longitudes from $l>325^\circ$
to $l<50^\circ$ (from the Norma spiral arm tangent to the Sagittarius spiral
arm tangent). These data allowed us to obtain the time averaged map of the
region and to construct a flux limited sample of sources with a limiting
sensitivity of $\sim$1.5 mCrab in the 20-60 keV energy band, that
corresponds to an energy flux $\sim1.8\times 10^{-11}$ ergs s$^{-1}$
cm$^{-2}$ for a Crab-like spectrum. This criterion was chosen to provide a
maximal survey area with best possible sensitivity.

%=====================================================================
\begin{table*}[t]
\caption{List of HMXB located in the Galactic plane from the Norma to 
Sagittarius spiral arms which were detected with INTEGRAL}\label{list}
\begin{center}
\begin{tabular}{lrrcccl}
\hline
\hline
Source & $\alpha$(2000) & $\delta$(2000) &N$_{\rm H}$,$10^{22}$
cm$^{-2}$ &N$_{\rm H}$,$10^{22}$
cm$^{-2}$& Flux, mCrab\a & Comments\b\\
&&&Observed&Galactic&&\\
\hline
4U 1538-522    & 235.584& -52.376  &  1.6[1] & 0.96 &  16.5$\pm$0.2   &  P \\    
AXJ161929-4945 & 244.871& -49.758  &  14\c  & 2.19 &2.0$\pm$0.2    &  \\
IGR J16318-4848& 247.953& -48.801  &  310\c&  2.07 &21.8$\pm$0.2   &  \\
AX J1631.9-4752\d& 248.009& -47.859&  18\c&  2.13 &13.0$\pm$0.2   &  P\\
IGR J16358-4726& 248.990& -47.407  &  40\c&  2.20 &3.11$\pm$0.21  &  P \\
AX J163904-4642& 249.757& -46.676  &  58\c&  2.18 &4.63$\pm$0.21  &  P \\
IGR J16465-4507& 251.648& -45.118  &  72\c&  2.12 &8.8$\pm$0.9    &  P  \\   
IGR J16479-4514\e& 252.032& -45.206&  12\c&  2.14 &3.22$\pm$0.20  &  \\
OAO 1657-415   & 255.199& -41.653  &  40[2]&  1.76 &68.2$\pm$0.2   &  P \\
4U 1700-377    & 255.982& -37.841  &  2-100[3] &  0.74 &185.0$\pm$0.2  &  \\
EXO 1722-363   & 261.286& -36.280  &  50[4]  &  1.50 &7.18$\pm$0.13  &  P \\
IGR/XTE J17391-3021& 264.802& -30.329& 5-6[5] &  1.37 &1.50$\pm$0.11  &  \\
AX J1749.2-2725& 267.335& -27.511  &  10[6]  &  1.62 &1.51$\pm$0.11  &  P \\    
IGR/SAX J18027-2016& 270.677& -20.278&  1-1.5[7] &  1.04 &4.06$\pm$0.14  &  P \\
AX J1820.5-1434& 275.131& -14.553  &  13[8]  &  1.65 &2.94$\pm$0.20  &  P\\
AX J1838.0-0655& 279.523&  -6.921  &   9\c  &  1.86 &2.36$\pm$0.24  &  \\
GS 1843+00     & 281.412&  0.891   &   2.3[9]  &  1.01 &4.49$\pm$0.20  &  P \\
XTE J1855-026  & 283.873& -2.597   &  15[10]  &  0.73 &11.4$\pm$0.2   &  P \\
4U 1901+03     & 285.914& 3.215    &   7\f  &  1.03 &75.5$\pm$0.2   &  P \\
4U 1907+097    & 287.401& 9.843    &  3-8 [11] &  1.75 &13.2$\pm$0.2   &  P \\
X1908+075      & 287.699& 7.598    &  10-50[12] &  1.48 &13.4$\pm$0.2   &  P\\  
SS 433         & 287.950& 4.990    &$-$\g &  0.76 &14.2$\pm$0.2   &  BH \\
XTE J1858+034  & 284.693& 3.429    &   6 [13] &  1.89 &15$\pm$1     &  P \\
\hline
 
\end{tabular}
\end{center}
\vspace{2mm}

\begin{list}{}
\item $^a$ -- in the 20-60 keV energy band;
\item $^b$ -- X-ray pulsar (P), black-hole candidate (BH);
\item $^c$ -- photoabsorption column obtained from our analysis of the broadband spectrum (see Table 2) 
\item $^d$ -- the same as IGR J16320-4751;
\item $^e$ -- source localization was improved in our analysis of the 
archival data of the ASCA observatory:  ${\rm RA}= 16^h 48^m07^s$, 
${\rm Dec}= -45$\deg$12$\arcmin$21$\arcsec (an accuracy of 0.9\arcmin);
\item $^f$ -- determined from the analysis of the public RXTE data;  
\item $^g$ -- central X-ray source is completely absorbed; the observed X-ray 
emission emerges from a hot outflowing jet (e.g.\cite{watson86});
\item References: [1] --  \cite{robba01}, [2] -- \cite{chakrabarti02}, 
[3] -- \cite{haberl92}, [4] -- \cite{lut04a}, [5] -- \cite{smith98},
[6] -- \cite{torii98}, [7] -- \cite{augello03}, [8] -- \cite{kinugasa98}, [9] -- \cite{piraino00}, [10] -- \cite{corbet99}, [11] -- \cite{roberts01}, [12] -- \cite{levine04}, [13] -- \cite{paul98}.
\end{list}

\end{table*}
%=====================================================================
%=====================================================================
\begin{figure}[b]
\includegraphics[width=\columnwidth,bb=56 186 565 424,clip]{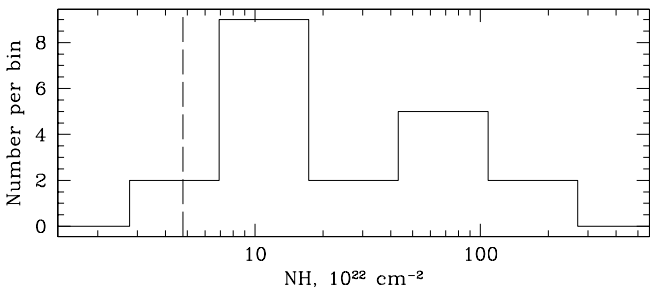}
\caption{Distribution of the absorption column value observed in 
X-ray spectra of HMXBs of our sample. The dashed line denotes the area
where systematic uncertainties caused by high interstellar absorption
in some Galactic regions prevents accurate measurements the source
intrinsic photoabsorption value.}
\label{nhdist}
\end{figure}
%=====================================================================

In the selected region of the Galactic plane ($325^\circ<l<50^\circ$)
INTEGRAL detects 89 different sources with a time average flux of $\ga$1.5
mCrab in the 20-60 keV energy band. Among these sources there are 49
identified low mass X-ray binaries, 22 high mass X-ray binaries and 2 active
galactic nuclei. Also there are supernova remnants and single pulsars. Ten
sources remain unidentified. A detailed analysis of the available INTEGRAL
sky map will be presented in a separate paper.

The list of high mass X-ray binaries presented in our sample is shown in
Table~\ref{list}. In this Table we also present the value of the observed
X-ray photoabsorption (column 3), the value of the interstellar Galactic
$N_{\rm H}$ value inferred from radio maps (\cite{nh}), the average source
flux in mCrab values and comments about the source nature.

Most of the presented HMXBs (16 of 23) are accretion-powered X-ray pulsars,
one is a black hole candidate, 4U1700-377 is a well-known high mass X-ray
binary without detectable pulsations. Other sources were included in the
list because of the strong intrinsic absorption in their X-ray spectra that
is not typical for low mass X-ray binaries (IGR/XTE J17391-3021 was included
based on the identification of the compaion star as a late O-type star,
\cite{smi04}). One of the sources, IGR J16465-4507, was not detected in the
INTEGRAL time averaged map, as it was discovered only in September 2004
(revolution 232, \cite{lut04b}). Follow-up observations with the XMM-Newton
observatory revealed a strong absorption in the source spectrum (see
Table~\ref{list}). Moreover, the improvement of the source position
(\cite{zur04}) allowed observers to propose a blue supergiant star as its
possible counterpart in the system (\cite{smi04}). The timing analysis of
XMM-Newton data allowed us to detect pulsations of X-ray flux from this
source (see below). Therefore we included IGR J16465-4507 in the list of
HMXBs and have a sample of 23 sources in total.
%=====================================================================
\begin{table*}[t]
\caption{Best-fit spectra parameters of 8 highly absorbed sources in the 
1-100 keV energy band}\label{spcpar}
\begin{center}
\begin{tabular}{lcccc}
\hline
\hline
Source & $\Gamma$ & N$_H$,$10^{22}$ cm$^{-2}$& $E_{c}$, keV & $\chi^2$/(d.o.f)\\ 
\hline
AX J161929-4945& 1.3$\pm$1.1 & 14$\pm$4 &   31$\pm$8  & 0.64 \\
IGR J16318-4848& 1.0$\pm$0.5 &310$\pm$70&   22$\pm$1  & 0.95 \\
AX J1631.9-4752& 0.7$\pm$0.2 & 18$\pm$2 &   13$\pm$1  & 1.17 \\
IGR J16358-4726& 0.7$\pm$0.5 & 40$\pm$10&   16$\pm$5  & 1.16 \\
AX J163904-4642& 1.3$\pm$1.0 & 58$\pm$11&   11$\pm$1  & 0.57 \\
IGR J16465-4507& 1.0$\pm$0.5 & 72$\pm$6 &   30        & 0.68 \\   
IGR J16479-4514& 1.4$\pm$0.8 & 12$\pm$4 &   32$\pm$2  & 0.97 \\
AX J1838.0-0655& 1.5$\pm$0.5 &  9$\pm$3 &   $>50$     & 0.82 \\
\hline
\end{tabular}
\end{center}
\end{table*}
%=====================================================================  

We can estimate the number of extragalactic (mostly AGN) sources we should
detect in the studied region with our sensitivity limit.  For this purpose
we can use either the flux-number function of Krivonos et al. (2005) in the
energy band 20-60 keV or use the more conventional flux-number function in
the 2-10 keV energy band (e.g. \cite{ueda03,sazonov04}) with the rescaled
sensitivity limit (assuming, for example, a photon index of AGN spectra of
$\Gamma\sim 1.7$). In both cases the estimated number of AGN in our region
is of the order of 3-4 sources. Two AGN have already been identified -- GRS
1734-292 (\cite{pav92}) and IGR18027-1455 (\cite{mas04}).

\section{Absorbed sources}

The study of absorbed X-ray sources attracted much of attention with the
launch of INTEGRAL (\cite{matt03}, \cite{rev03}, \cite{rod03}, \cite{wal03},
Foschini et al. 2004, \cite{com04}, etc.). INTEGRAL is able to detect and
measure the spectra of sources which are practically invisible for low
energy X-ray instruments. The population of sources that demonstrates a high
value of photoabsorption is not so small as to be disregarded -- about 19
sources out of 23, presented in Table 1 have a significant ($>5\times
10^{22}$ cm$^{-2}$) intrinsic photoabsorption.  In Fig.~\ref{nhdist} we
present the distribution of the observed $N_{\rm H}$ values in spectra of
sources in our sample.

\subsection{Spectra}

\begin{figure*}[p]
\hbox{
\includegraphics[width=\columnwidth,bb=20 410 555 730]{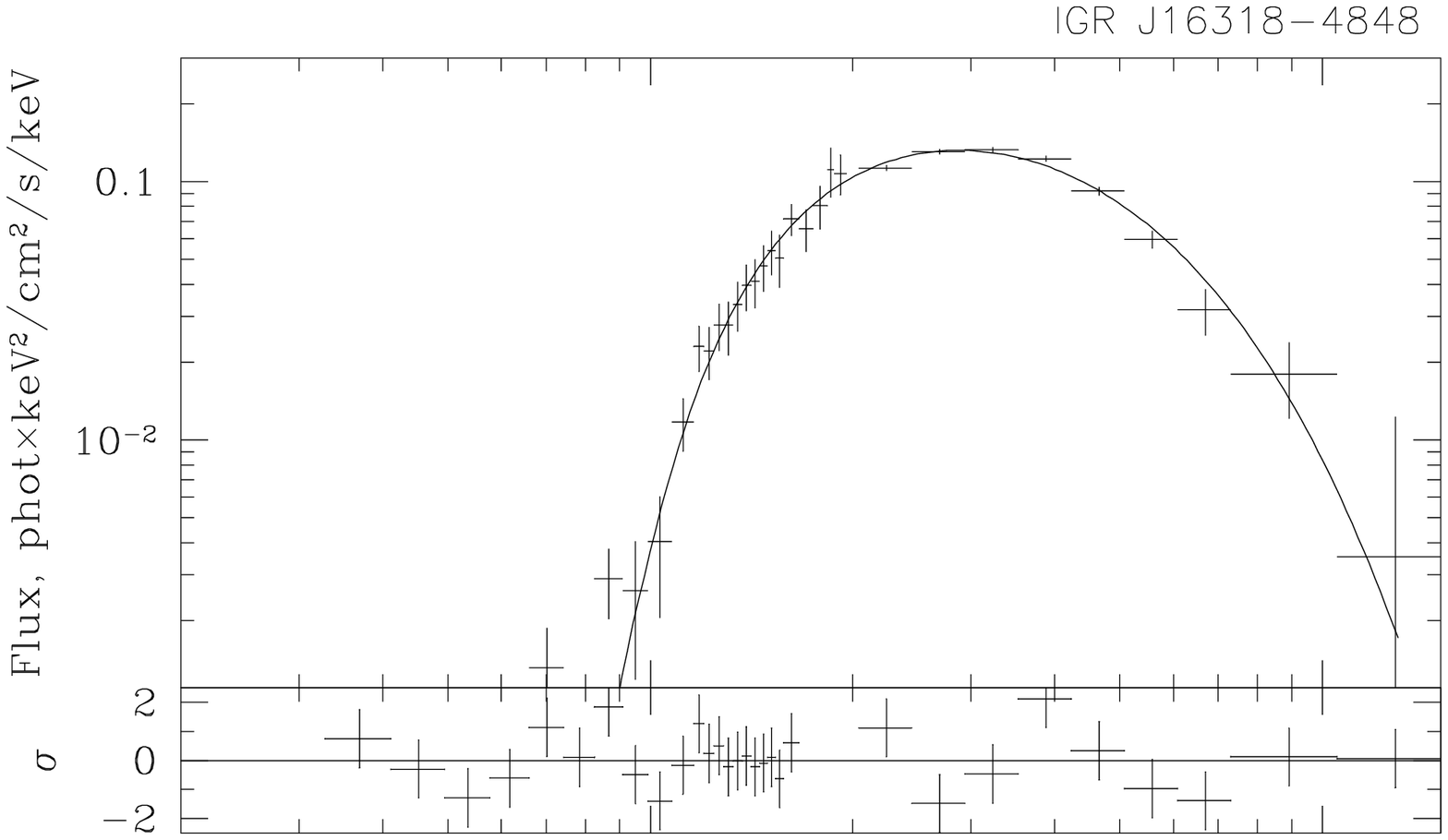} 
\includegraphics[width=\columnwidth,bb=40 410 575 730]{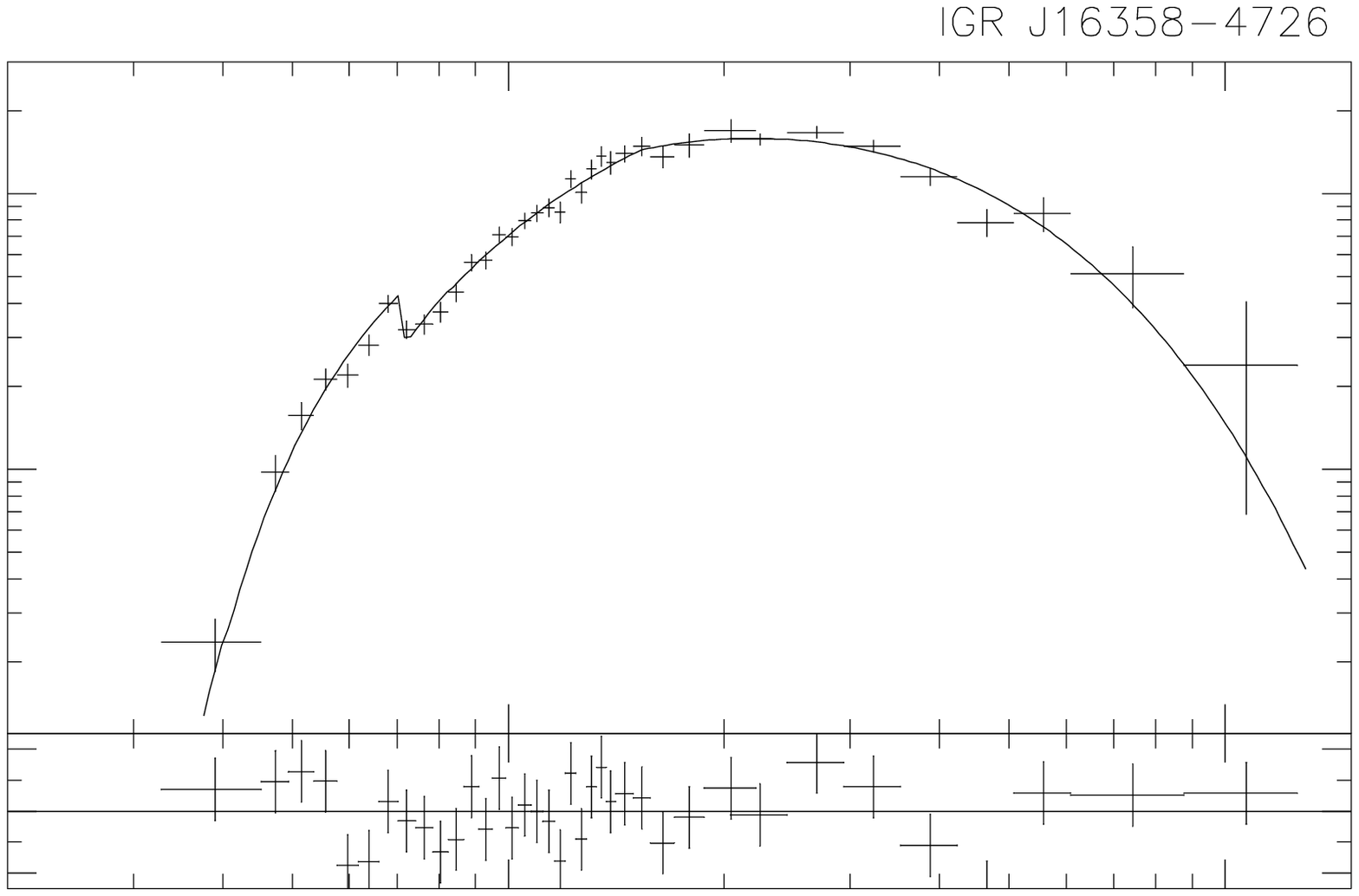}
}
\hbox{
\includegraphics[width=\columnwidth,bb=20 410 555 730]{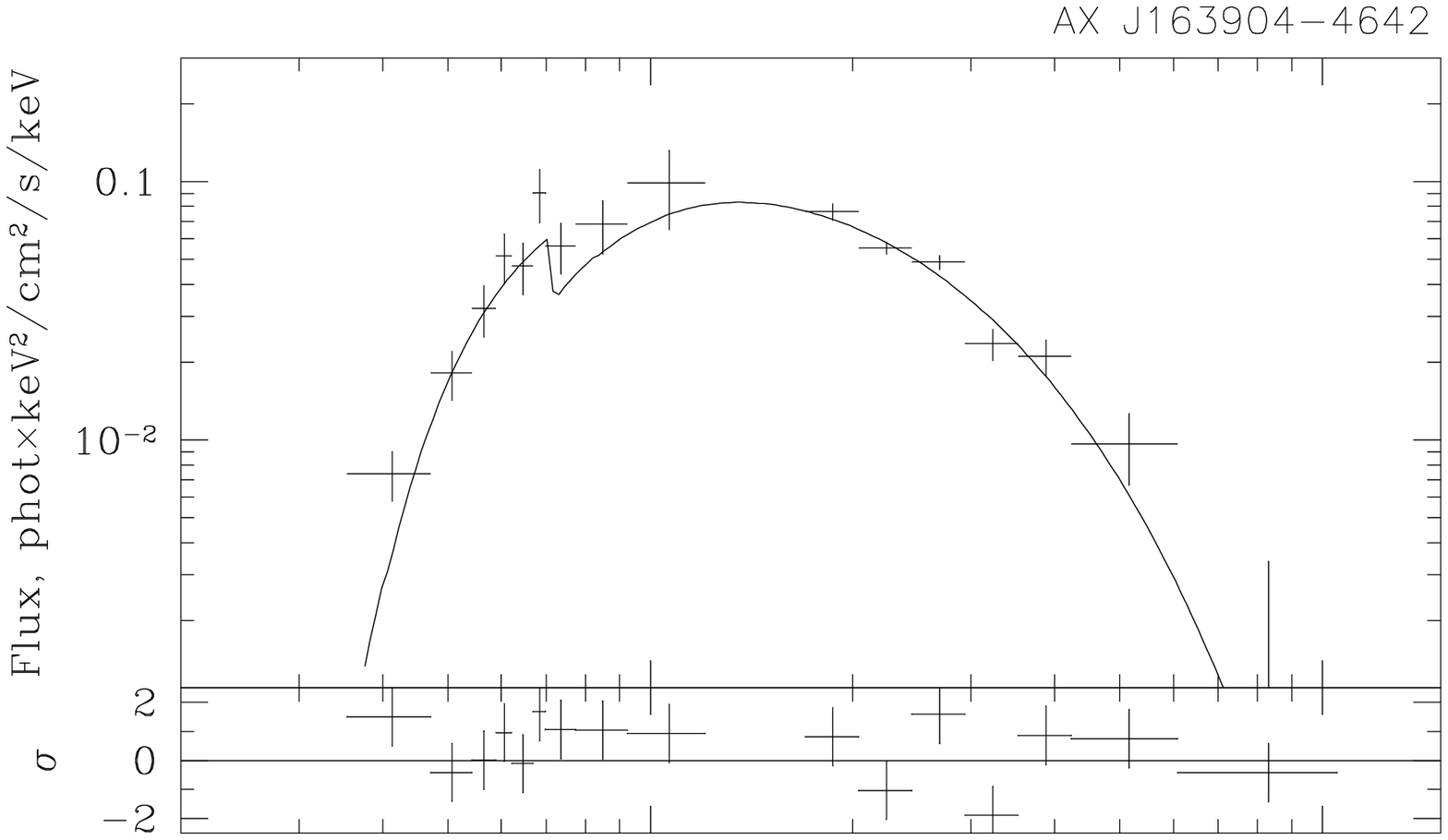}
\includegraphics[width=\columnwidth,bb=40 410 575 730]{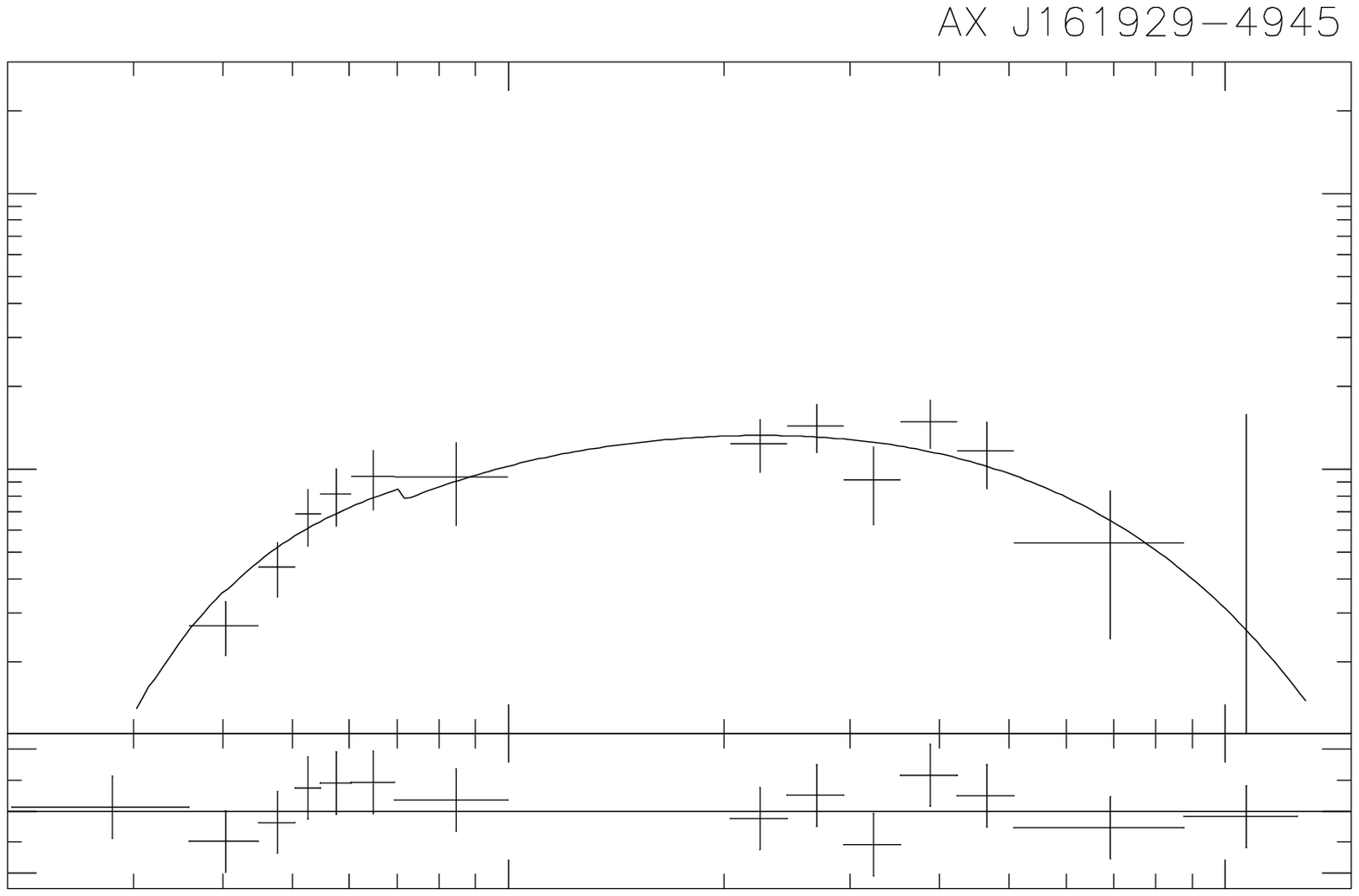}
}
\hbox{
\includegraphics[width=\columnwidth,bb=20 410 555 730]{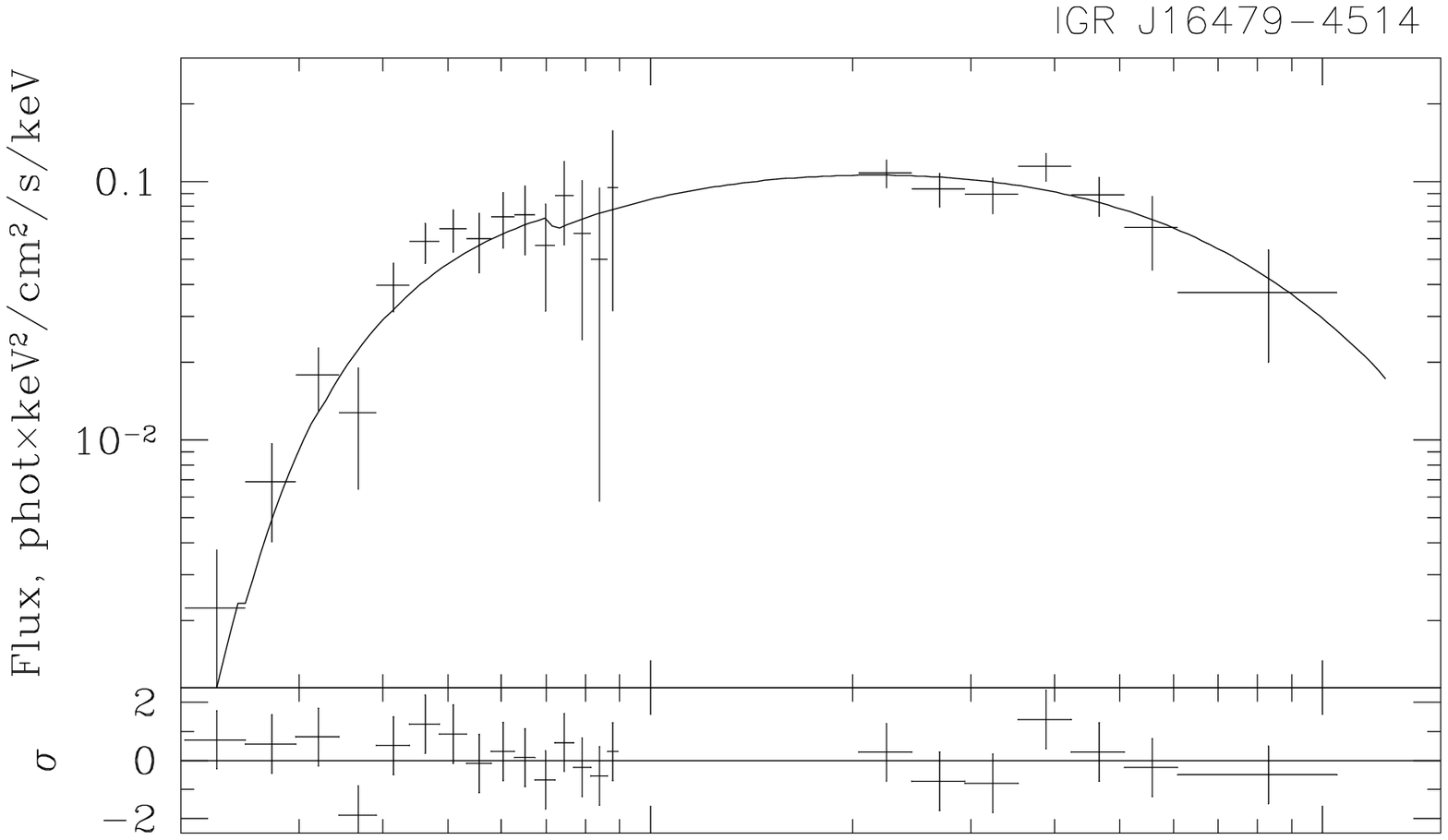}
\includegraphics[width=\columnwidth,bb=40 410 575 730]{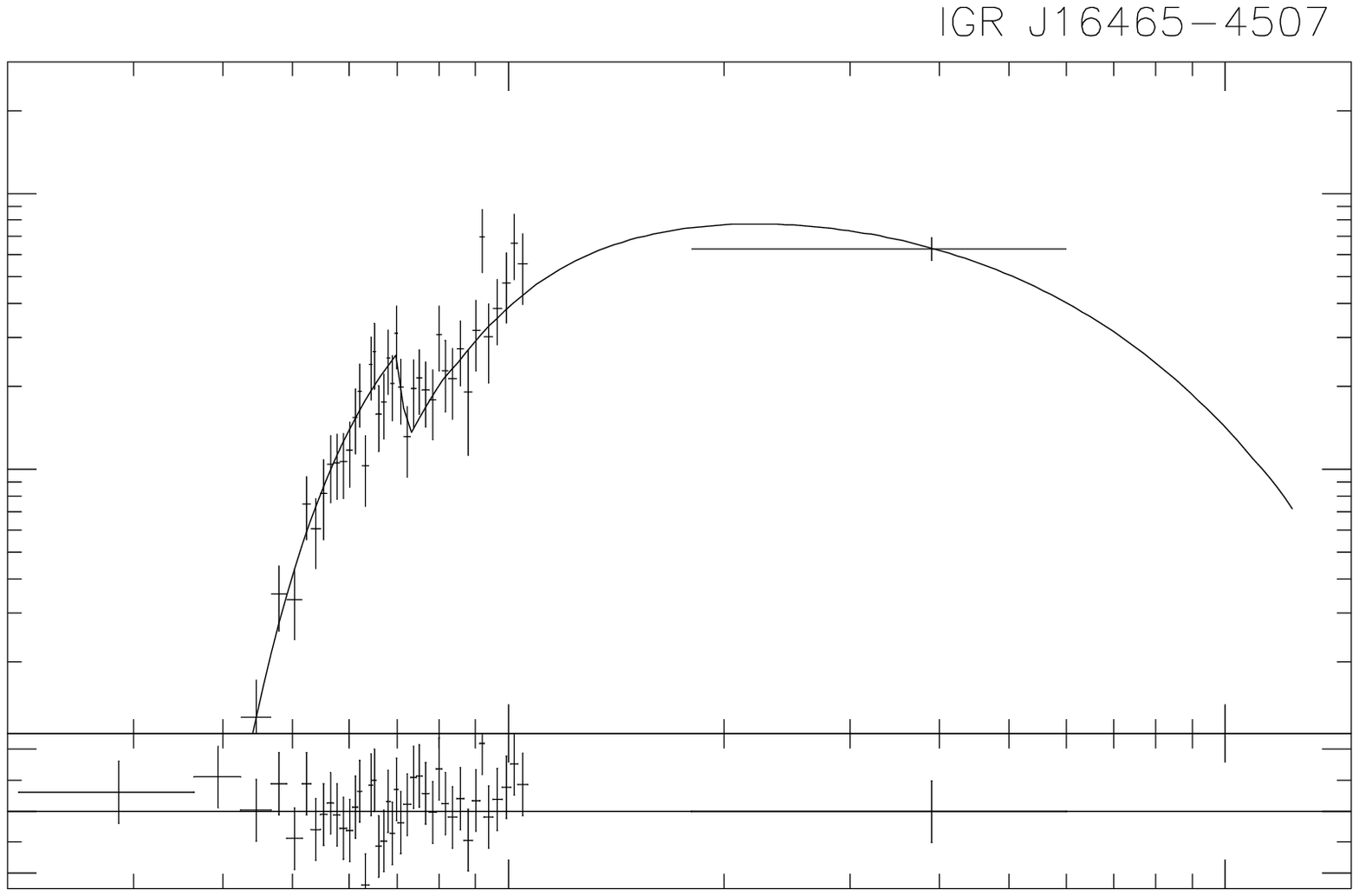}
}
\hbox{
\includegraphics[width=\columnwidth,bb=20 340 555 730]{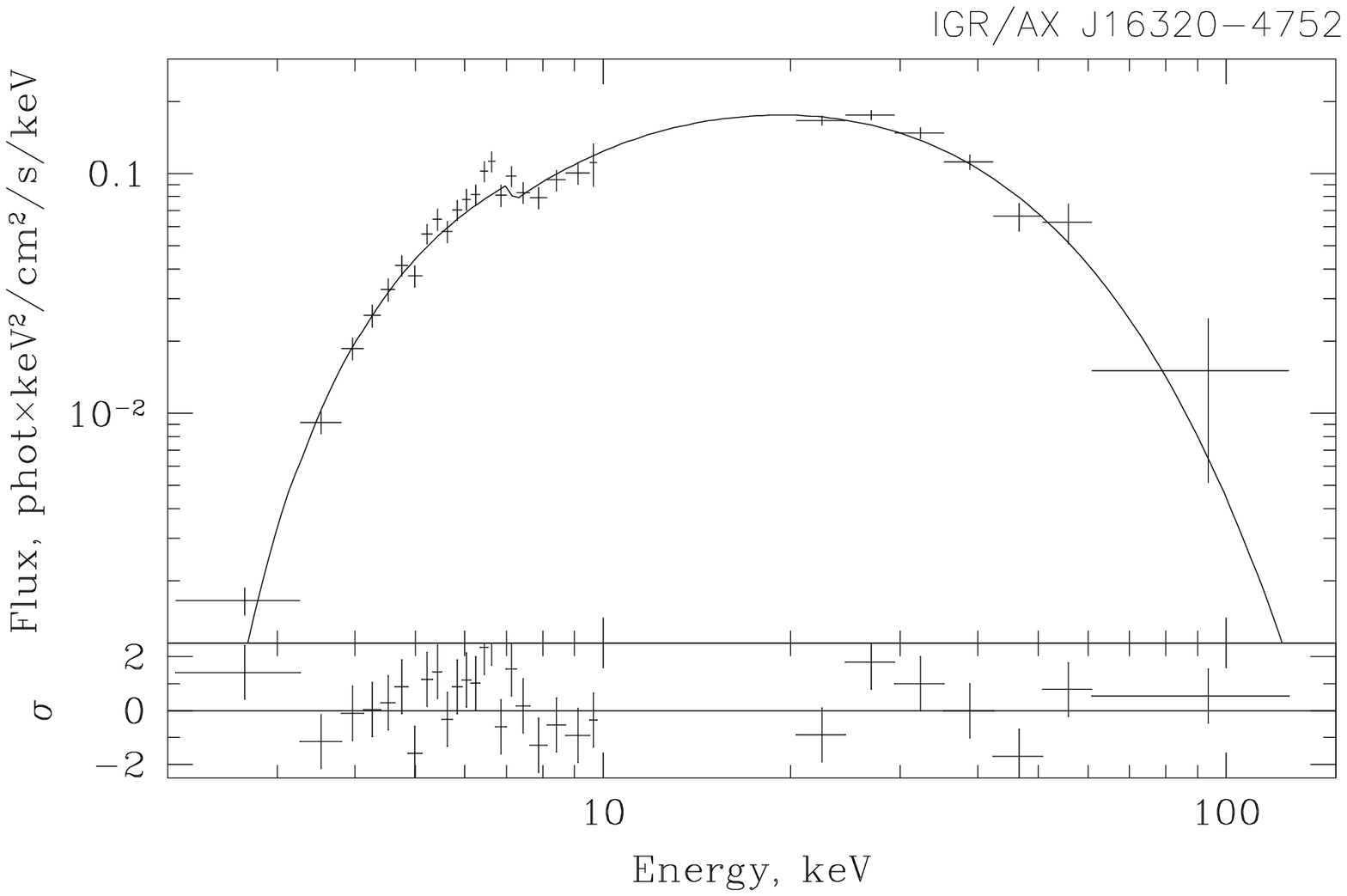}
\includegraphics[width=\columnwidth,bb=40 340 575 730]{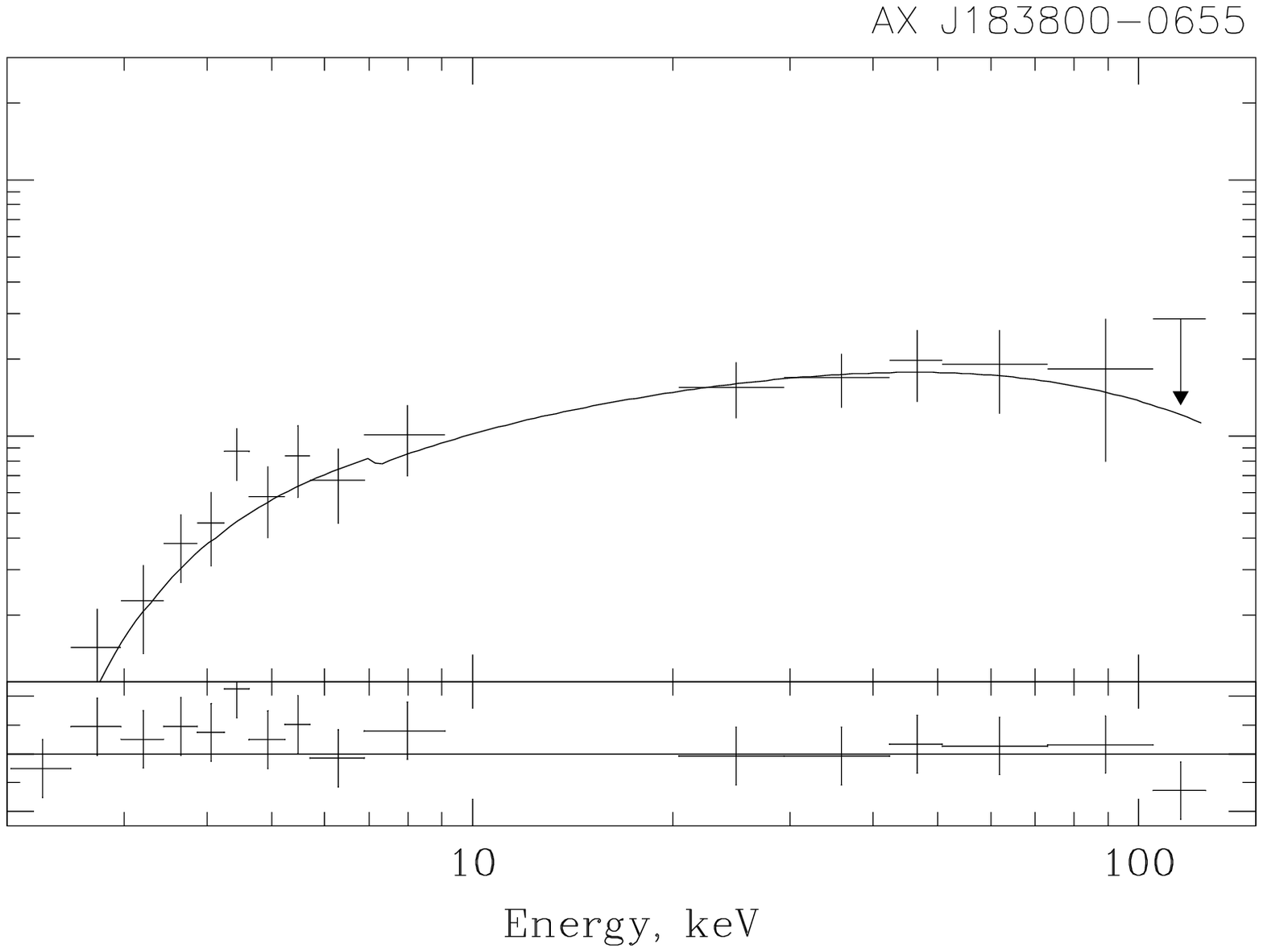}
}
\caption{Broadband energy spectra of 8 highly absorbed sources. Data in the 
hard energy part of spectra ($>20$ keV) represent the INTEGRAL observatory
measurements; data in standard X-rays were obtained by ASCA for AX
J161929-4945, AX J1631.9-4752, AX J163904-4642, IGR J16479-4514, AX
J1838.0-0655, RXTE for IGR J16318-4848, IGR J16358-4726 and XMM-Newton for
IGR J16465-4507. Best fit models are shown by solid lines. Residuals are
presented in the bottom panels.} \label{spectra}
\end{figure*}

To reconstruct the broadband spectra of absorbed sources we combine all
available data from ASCA, RXTE, XMM-Newton and INTEGRAL observatories. The
energy spectra of 8 sources in the 1--100 keV energy band are presented in
Fig.~\ref{spectra}. These sources are the least studied in the broad energy
band binaries located in the inner part of the Galaxy; most of them were
discovered or rediscovered with INTEGRAL in hard ($>20$ keV) X-rays.
Spectral points in a soft energy band ($<20$ keV) obtained with
non-simultaneous observations of ASCA, XMM-Newton and RXTE were renormalized
to match the INTEGRAL/IBIS spectral normalization. Here we assumed that the
shape of the source spectra is not variable and used time averaged data.

The obtained spectra can be well described by a simple powerlaw model
modified by the cutoff at high energies and photoabsorption at soft X-rays
that is typical for accreting neutron stars (see e.g. the spectrum of
IGR/XTE J17391-3021 observed with INTEGRAL in the Galactic Center region,
\cite{lut05b}). Best fit parameters are summarized in
Table~\ref{spcpar}. Best fit models are shown in Fig.~\ref{spectra} by solid
lines.

Note that for the source IGR J16465-4507 the spectrum is based on the data
of XMM-Newton observations performed on Sept.14, 2004 and the measurement of
INTEGRAL (\cite{lut04b}). The solid line represents the spectral form
typical of new absorbed sources -- a power law with an exponential cutoff
(this value was fixed at the energy $E_{\rm c}=30$ keV to better approximate
the INTEGRAL point) and modified with the photoabsorption, determined with
the help of XMM-Newton data.

It is seen that most of the sources have a somewhat similar spectral shape
and parameters, in particular the folding (cutoff) energy. Only one source
(AX J18380-0655) demonstrates a high energy cutoff value in its spectrum
above $\sim 30$ keV. Such high values of the cutoff are not typical for
neutron star systems, and therefore this sources is likely a black hole
candidate. All other systems demonstrate X-ray spectra typical of accreting
neutron stars.

%===================================================================
%===================================================================

\subsection{Pulsations in IGR J16358-4726 and IGR J16465-4507}

For some of the newly discovered sources coherent pulsations of the X-ray
flux were detected -- IGR J16358-4726 (\cite{patel04}), AX J163904-4642
(\cite{walter04}), AX J1631.9-4752 (\cite{lut05a}).

The available data of INTEGRAL and XMM-Newton observatories allow us to
confirm pulsations in the flux of IGR J16358-4726 and detect pulsations in
the flux of IGR J16465-4507.

%=====================================================================  
\begin{figure}[t]
\includegraphics[width=\columnwidth,bb=30 350 565 695,clip]{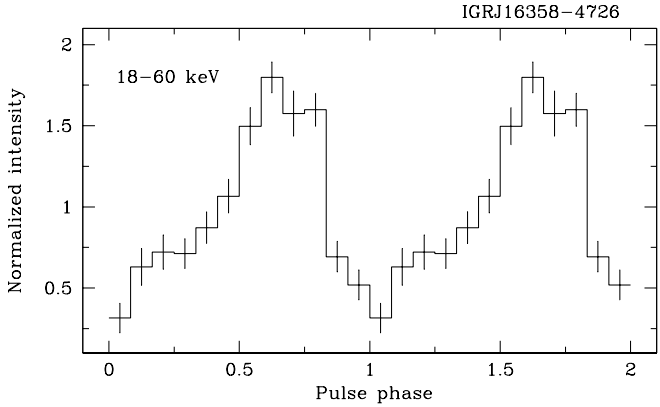}
\caption{Folded light curve of IGR J16358-4726 obtained with 
INTEGRAL/IBIS/ISGRI in the 18-60 keV energy band}\label{16358pp}
\end{figure}
%=====================================================================  

INTEGRAL observed IGR J16358-4726 during its outburst in spring of 2003
(\cite{rev03b,rev03c}). The source lightcurve in the 18-60 keV energy band
demonstrates clear pulse-like variations. The total exposure time of
INTEGRAL observations used to construct the light curve of IGR J16358-4726
at the peak of its outburst was about $\sim90$ ksec. The best period,
obtained with the epoch-folding technique, is $P=5980\pm22$ sec (the error
represents the 90\% confidence interval; it was determined from the
bootstrap Monte-Carlo simulations). The pulse profile of IGR J16358-4726 is
presented in Fig.~\ref{16358pp}.  As in the softer energies (see
\cite{patel04}) the source pulse profile consists of one peak. The pulse
fraction in the 18-60 keV energy band (which is defined as
$P=(I_{max}-I_{min})/(I_{max}+I_{min})$, where $I_{max}$ and $I_{min}$ are
intensities at the maximum and minimum of the pulse profile) is equal to
$P=70\pm10$\%, the same as in the CHANDRA energy band.  The detailed timing
analysis of the data of the XMM-Newton observations of IGR J16465-4507
allowed us to increase the number of known accreting X-ray pulsars in the
region of the Norma spiral arm.
%=====================================================================
\begin{figure}[t] \includegraphics[width=\columnwidth,bb=30 120 570
700,clip]{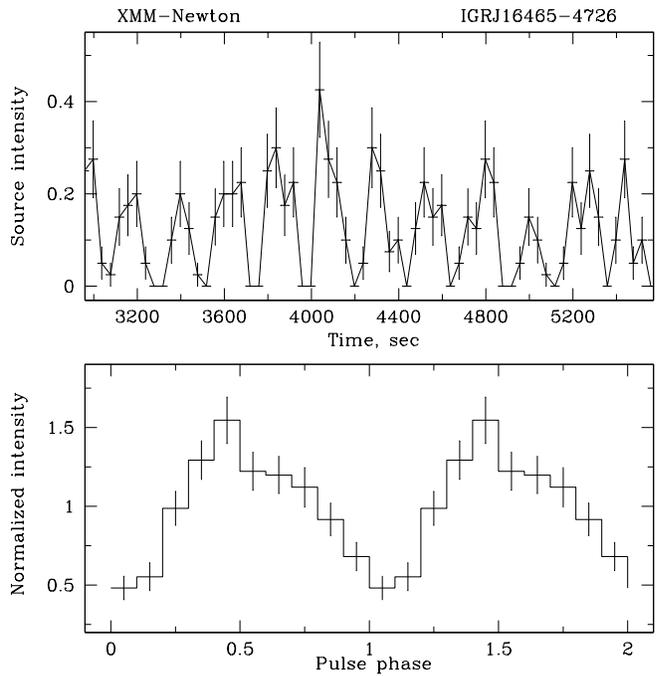} \caption{({\it upper panel}) Part of the
light curve of IGR J16465-4507 obtained by XMM-Newton (the time bin is 40
sec). ({\it bottom panel}) Corresponding pulse profile of IGR J16465-4507
folded with the period of 228 sec. } \label{16465pp} \end{figure}
%=====================================================================

The light curve of IGR J16465-4507 obtained with the XMM-Newton observatory
during TOO observations of the source on Sept. 14, 2004 is shown in
Fig.~\ref{16465pp}. Strong variations are presented in the light
curve. The best fit period determined with the epoch-folding technique is
$P=228\pm6$ sec; the pulse fraction was estimated as $\sim50-60$\%. The
folded light curve is shown in Fig.~\ref{16465pp} (lower panel). Note that
zero flux bins in the original light curve are caused by the Poissonian
statistics. The mean count rate during the pulsation minimum is $\sim$ 0.03
cnts/s, or $\sim 1$ cnts per bin (40 sec). We could not search for
pulsations in the INTEGRAL data because they are currently not publically
available.

For other sources no pulsations have been detected yet. However,
similarities of their spectral shape to that of confirmed high mass X-ray
binaries and the presence of the strong absorption in their spectra allow us
to tentatively suggest that all they are HMXBs.

\section{Spatial distribution of the X-ray binaries}

The obtained sample of Galactic X-ray binaries allows us to study their
spatial distribution in the comparison with the previous work of Grimm et
al. (2002). In that work authors used the data of RXTE/ASM (energy band 1-12
keV) and that less sensitive to sources with high intrinsic
photoabsorption. In addition to that, our limiting sensitivity is slightly
better than that of Grimm et al. (2002). Assuming a spectrum of a typical
accreting X-ray pulsar with zero intrinsic photoabsorption (power law with
photon index $\Gamma=1.3$ with a high energy cutoff at 30 keV) our
sensitivity of $\sim1.8\times 10^{-11}$ ergs s$^{-1}$ cm$^{-2}$ in the 20-60
keV energy band corresponds to $\sim 1-2\times 10^{-11}$ ergs s$^{-1}$
cm$^{-2}$ in the RXTE/ASM energy band 1-12 keV (cf. $\sim 6\times 10^{-11}$
ergs s$^{-1}$ cm$^{-2}$ in Grimm et al.  2002).

In the region of our study ($325^\circ<l<50^\circ$) we have detected in
total 23 high mass X-ray binaries and HMXB candidates. This is more than 2
times larger than the sample used by Grimm et al. (2002) in the same region.
In the region $l\sim340^\circ$ our sample contains 6 sources, while
Grimm et al. (2002) considered only one source. This is not a surprise
because a significant part of our sample consists of new sources discovered
by INTEGRAL which have strong photoabsorption. Note however, that the
total sample of HMXBs used by Grimm et al.(2002) includes more than 50 HMXBs
and is larger than ours.

%=====================================================================  
\begin{figure}[]
\includegraphics[width=\columnwidth,bb=56 186 585 694,clip]{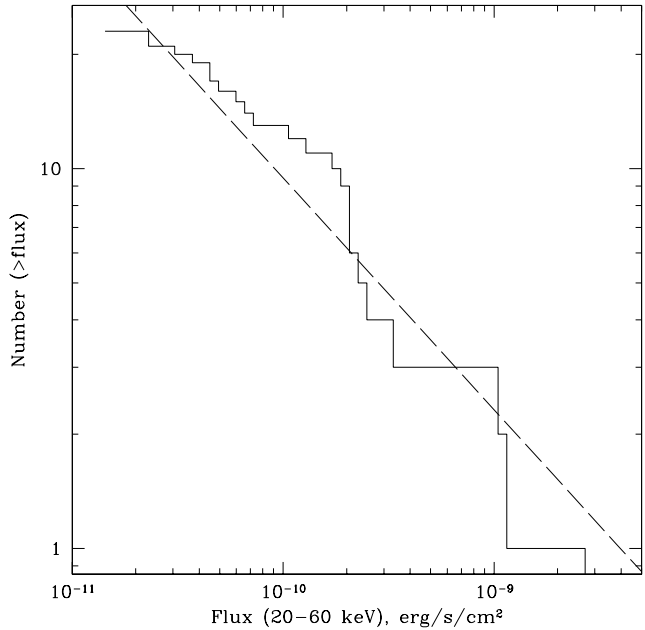}
\caption{Number-flux function of the Galactic HMXBs detected by 
INTEGRAL/IBIS/ISGRI in the inner part of the Galaxy. The dashed line
represents the shape of the HMXB number-flux function obtained by
Grimm et al. (2002)}
\label{logn}
\end{figure}
%=====================================================================  

The number-flux distribution of HMXB detected with INTEGRAL is presented in
Fig.~\ref{logn}. The dashed line represents the number-flux relation
obtained by Grimm et al. (2002).

%=====================================================================  
\begin{figure}[b]
\includegraphics[width=\columnwidth,bb=47 346 570 680,clip]{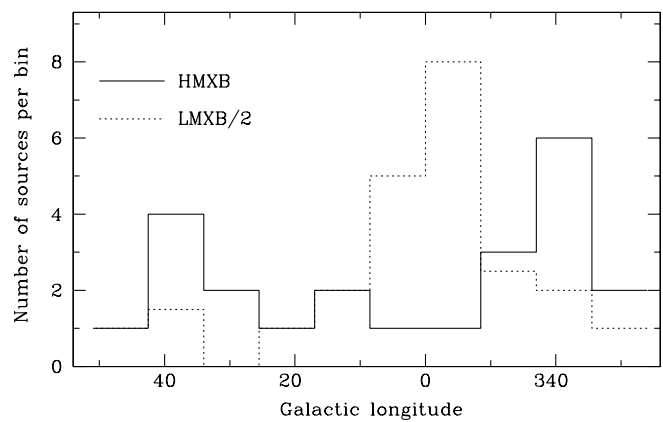}
\caption{Angular distribution of high mass X-ray binaries (HMXBs) and low 
mass X-ray binaries (LMXBs) in the inner Galaxy. The number of LMXB is
devided by 2.}
\label{hist_xb}
\end{figure}
%===================================================================== 

The angular distribution of high mass X-ray binaries along the Galactic
plane is presented by a histogram in Fig.~\ref{hist_xb} in the comparison
with the distribution of LMXBs. To estimate the significance of the
difference of the observed HMXB distribution from the LMXB one we used a
Kolmogorov-Smirnov test in two ways: for the whole sample
($325^\circ<l<50^\circ$) and for the first ($0^\circ<l<50^\circ$) and fourth
($325^\circ<l<360^\circ$) Galactic quadrants separately. We made this
division in order to put the regions of interest in the middle of trial
interval, where the K-S test is more sensitive. We build cumulative
distributions for each of these cases and found that the probabilities that
the HMXB distribution differs from LMXB one are $\sim 75$\% for the whole
sample and $\sim 96$\% and $\sim97$\% for first and fourth quadrants,
respectively (the significance of the central peak for the LMXB
distribution is much higher, $>99.9$\%). It is known (Grimm et al. 2002)
that low mass X-ray binaries trace the stellar mass of the Galaxy and are
concentrated in the Galactic center. From Fig.~\ref{hist_xb} and the K-S
test it follows that the high mass X-ray binary distribution is different to
the LMXB one and not concentrated at the Galactic center. HMXBs, being the
young population of X-ray sources in the Galaxy should be related to
star formation regions (e.g. Galactic spiral arms) rather than to the region
with high stellar mass density, and such a connection was indeed observed
(e.g. Grimm et al. 2002, Grimm et al. 2003).

%===================================================================== 
\begin{figure}[t]
\includegraphics[height=\columnwidth]{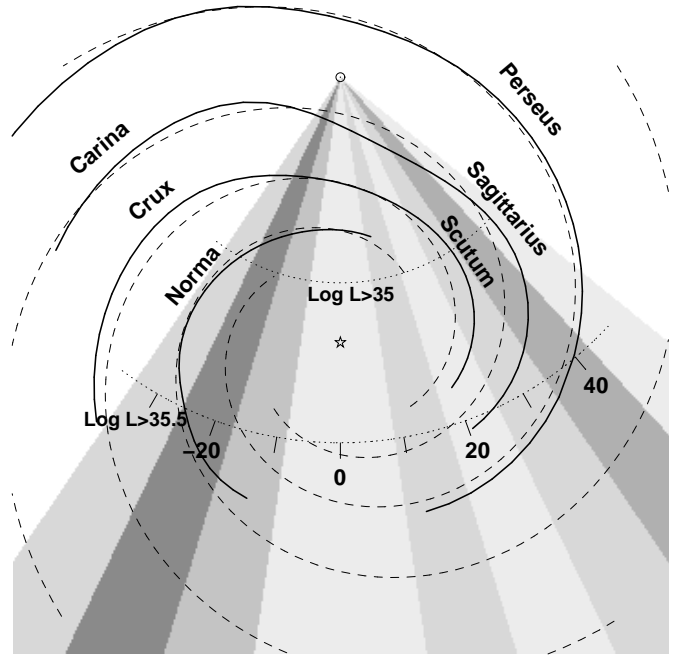}
\caption{Face-on view of the Galaxy with overlayed densities of HMXBs 
shown in gray scale (from 1 to 6, see Fig.~\ref{hist_xb}). 
The distance of 8.5
kpc of the Galactic Center from the Sun is assumed. The star denotes the
position of the Galactic Center. The open circle with the dot denotes the
position of the Sun. Solid lines show the position of the spiral arms
by Taylor \& Cordes (1993), the dashed line shows logarithmic spiral arms
by Vallee (1995), see also Russeil (2003)}
\label{faceon}
\end{figure}
%===================================================================== 

In Fig.~\ref{faceon} we visualize the distribution of HMXB in the Galaxy
with respect to the Galactic spiral structure. The number density of the
sources is shown in gray scale. The spiral model of the Galaxy is based on
optical and radio observations of HII regions (\cite{taylor93}). The dotted
circles denote arcs of circles centered on the Sun and show maximal
distances at which our sensitivity limit ($1.5$ mCrab) allows us to detect
X-ray sources with the marked luminosities ($L_{\rm x}=10^{35}$ erg/s and
$L_{\rm x}=10^{35.5}$ erg/s). The dashed lines denotes four logarithm
spirals with pitch angle $12^\circ$ adopted by Vallee (1995).  This model
predicts the position of spiral arms in the opposite (with respect to the
Galactic Center) part of the Galaxy, where measurements in the optical and
radio bands are scarce, but these parts can be visible in X-rays.

The detailed comparison of the density of the high mass X-ray binary
distribution with the spiral structure of the Galaxy (regions of the
enhanced star formation) is not possible at the moment due to the limiting
statistics of the sample used. This is the subject of a separate paper in
which more INTEGRAL data for the whole Galaxy will be used. Here we only
mention several possible reasons that may affect the observed HMXB
density. They are: unknown distances to the sources and, as a sequence,
their unknown exact positions; dynamics of the Galaxy and complicated
motions of the Galactic spiral structure and stars from the moment of star
formation to the X-ray source stage (e.g. a significant displacement of
HMXBs with respect to the current position of the spiral arms was observed
by CHANDRA in spiral galaxy M83 \cite{soria03}); the possible influence of
previously unseen parts of the spiral arms, etc.

\section{Unidentified sources}

In order to understand the possible influence of unidentified sources in our
sample on the obtained distribution of HMXB we construct their distribution
along the Galactic plane with the same coordinate binning as in
Fig.~\ref{hist_xb}.

\begin{figure}[t]
\includegraphics[width=\columnwidth,bb=47 346 570 690,clip]{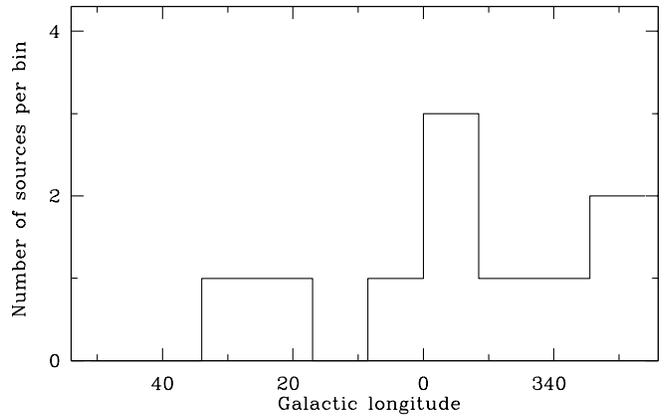}
\caption{The angular distribution of unidentified sources of our source sample along the Galactic plane.} 
\label{hist_un}
\end{figure}

This distribution (Fig.~\ref{hist_un}) shows that the unidentified sources
cannot affect our conclusion about the distribution of high mass X-ray
binaries in the inner part of the Galaxy. There is a small peak near the
Galactic Center that may indicate that several of the unidentified sources
are LMXBs.  One or two of the unidentified sources may be AGN.

%====================================================================
%====================================================================

\section{Summary}

\begin{itemize} 

\item We presented a sample of high mass X-ray binaries in the inner part of
the Galaxy from the Norma spiral arm region to the Sagittarius spiral arm
region ($-35^\circ<l<50^\circ$). This sample is significantly larger than
the HMXBs sample in the same region of the sky, based on RXTE/ASM data,
used by Grimm et al. (2002). The main reason for this is the hard X-ray energy
band (18--60 keV) used in the present observations which helped to reveal a
considerable population of absorbed HMXBs.

\item The majority of HMXB sources in our sample are significantly 
photoabsorbed.

\item We performed for the first time a spectral analysis for 8 strongly
absorbed sources of our sample. Based on the spectral approximations we
argue that the majority of these sources should be neutron star binaries
with a high mass companion. The exception is AX J183800-0655, which likely
contains a black hole.

\item We detected pulsations in the hard X-ray flux (18-60 keV) from IGR
J16358-4726 with the period $P=5980\pm22$ sec, thereby confirming the
results of the CHANDRA observatory (\cite{patel04}). In addition we
found pulsations with the period $P=228\pm6$ sec in the X-ray flux of IGR
J16465-4507 using the data of XMM-Newton.

\item We constructed the angular distribution of high mass X-ray binaries in
the inner part of the Galaxy and showed that this distribution differs from
the distribution of LMXBs, which are concentrated in the Galactic Center.

\item A large fraction of the accretion powered X-ray pulsars in the
Be-systems are transients. Therefore the continuation of INTEGRAL
observations of the Galactic plane might lead to the discovery of many more
similar systems and considerably increase the statistical significance of
the observed displacement of the HMXB population with respect to the spiral
arm tangents.  Data of CHANDRA and XMM-Newton on other galaxies and detailed
study of systems in our Galaxy might provide additional information about
the motion of stars and spiral waves.

\end{itemize}

\begin{acknowledgements} The authors thank Eugene Churazov for developing
methods of analysis of the IBIS data and software. We would like to thank
Sergei Sazonov for useful comments and the anonymous referee for critical
review which allowed us to significantly improve the paper. This research
has made use of data obtained through the INTEGRAL Science Data Center
(ISDC), Versoix, Switzerland, Russian INTEGRAL Science Data Center (RSDC),
Moscow, Russia and High Energy Astrophysics Science Archive Research Center
Online Service, provided by the NASA/Goddard Space Flight Center. This work
was partially supported by the program of Russian Academy of Sciences
``Non-stationary phenomena in astronomy''. AL, MR, PS and SM acknowledge the
support of RFFI grant 04-02-17276.

\end{acknowledgements}

\end{document}